\documentclass[twocolumn,prl,superscriptaddress]{revtex4}
\usepackage[utf8]{inputenc}
\usepackage{amsmath}
\usepackage{amssymb}
\usepackage{graphicx}
\usepackage{esint}
\usepackage{color}

\makeatletter
\@ifundefined{textcolor}{}
{%
 \definecolor{BLACK}{gray}{0}
 \definecolor{WHITE}{gray}{1}
 \definecolor{RED}{rgb}{1,0,0}
 \definecolor{GREEN}{rgb}{0,1,0}
 \definecolor{BLUE}{rgb}{0,0,1}
 \definecolor{CYAN}{cmyk}{1,0,0,0}
 \definecolor{MAGENTA}{cmyk}{0,1,0,0}
 \definecolor{YELLOW}{cmyk}{0,0,1,0}
}

\usepackage{subfigure}\usepackage{epsfig}\usepackage{amsfonts}\usepackage{mathrsfs}\usepackage{CJK}\usepackage[toc,page,title,titletoc,header]{appendix}

\makeatother

\begin{document}

\title{Calibration of the interaction energy between Bose and Fermi superfluids}

\author{Ren Zhang}

\affiliation{Department of Physics, Renmin University of China, Beijing, 100872,
China}

\affiliation{Institute for Advanced Study, Tsinghua University, Beijing, 100084,
China}

\author{Wei Zhang}

\affiliation{Department of Physics, Renmin University of China, Beijing, 100872,
China}

\affiliation{Beijing Key Laboratory of Opto-electronic Functional Materials \&
Micro-nano Devices, 100872 (Renmin Univeristy of China)}

\author{Hui Zhai}

\affiliation{Institute for Advanced Study, Tsinghua University, Beijing, 100084,
China}

\author{Peng Zhang}

\affiliation{Department of Physics, Renmin University of China, Beijing, 100872,
China}

\affiliation{Beijing Key Laboratory of Opto-electronic Functional Materials \&
Micro-nano Devices, 100872 (Renmin Univeristy of China)}
\begin{abstract}
In this paper we study the interaction energy in a mixture of Bose
and Fermi superfluids realized in recent cold atom experiment. On
the Bose-Einstein-condensate (BEC) side of a Feshbach resonance between
fermionic atoms, this interaction energy can be directly related to
the scattering length between a bosonic atom and a dimer composed
of fermions. We calculate the atom-dimer scattering length from a
three-body analysis with both a zero-range model and a separable model
including the van der Waals length scale, and we find significant
deviation from the result given by a mean-field approach. We also
find that the multiple scattering between atom and dimer can account
for such a deviation. Our results provide a calibration to the mean-field
interaction energy, which can be verified by measuring the shift of
collective oscillation frequency. 
\end{abstract}
\maketitle

\section{Introduction}

Few-body problems play many important roles in the study of cold atom
gases, partly due to the diluteness condition which is generally fulfilled
in the underlying systems. One of the most important examples is that
few-body results, which can be numerically exact, can provide benchmark
and calibration of many-body theories, where approximations are usually
inevitable. For instance, in the study of BEC-BSC crossover of a Fermi
superfluid around a Feshbach resonance, the mean-field theory yields
that the scattering length between fermion pairs (or dimers) equals
$2a_{\text{s}}$, with $a_{{\rm s}}$ the scattering length between
fermionic particles \cite{mean-field1,mean-field2}. However, a precise
four-body calculation gives a result of $0.6a_{\text{s}}$ \cite{few-body}.
This significant deviation suggests that it is necessary to include
pair fluctuations in order to obtain a more accurate many-body description.
Indeed, pair fluctuation theory can reduce this dimer scattering length
to $0.75a_{\text{s}}$ \cite{pair_flucuation}, which is much closer
to the few-body result. Moreover, experimental measurements of interaction
energy have confirmed the prediction from few-body calculation \cite{exp_dimer}.

Recently, the ENS Group has realized the first mixture of Bose and
Fermi superfluids with bosonic $^{7}$Li and two spin components of
$^{6}$Li \cite{exp}. To reach Fermi superfluid, the magnetic field
has to be tuned around a Feshbach resonance between two fermionic
components. In this regime all other interaction parameters are small.
For instance, in this experiment the boson-fermion scattering length
$a_{\text{bf}}$ is only $40.8a_{0}$ (where $a_{0}$ is the Bohr
radius). With such a weak interaction between bosons and fermions,
one may expect that the interaction energy between Bose and Fermi
superfluids can be obtained by mean-field theory quite accurately.
In fact, such a mean-field and hydrodynamic theory treatment have
been used in analyzing this system \cite{exp,Stringari,pu,jetp}.
The purpose of this work is to provide a calibration of this mean-field
theory from few-body calculation of atom-dimer scattering length.

Considering the situation that boson-fermion scattering length $a_{\text{bf}}$
is independent of fermionic spin species (as  in the current experiment),
boson-fermion interaction is described by 
\begin{equation}
\hat{V}=\frac{2\pi\hbar^{2}a_{\text{bf}}}{m_{\text{bf}}}\sum\limits _{\sigma}\int d^{3}{\bf r}\hat{b}^{\dag}({\bf r})\hat{b}({\bf r})\hat{c}_{\sigma}^{\dag}({\bf r})\hat{c}_{\sigma}({\bf r}),
\end{equation}
where $\hat{b}^{\dag}$ and $\hat{c}_{\sigma}^{\dag}$ are creation
operators for bosons and fermions with spin $\sigma=\uparrow$ or
$\downarrow$, respectively, and $m_{{\rm bf}}$ is the reduced mass
of the boson and fermion. With Hartree-Fock mean-field decomposition, $\langle\hat{b}^{\dag}\hat{b}\rangle=n_{\text{b}}$
and $\langle\hat{c}_{\sigma}^{\dag}\hat{c}_{\sigma}\rangle=n_{\text{f}}$,
this interaction energy density is naturally given by 
\begin{equation}
\mathcal{E}=\frac{2\pi\hbar^{2}a_{{\rm bf}}}{m_{{\rm bf}}}n_{{\rm b}}n_{{\rm f}}.\label{Ebf}
\end{equation}

On the other hand, on the BEC side of the resonance, the Fermi superfluid
can be viewed as a Bose condensate of dimers. The interaction energy
between Bose and Fermi superfluids can then be considered as the boson-dimer
interaction. We introduce $a_{\text{ad}}$ as the scattering length
between bosonic atoms and dimers. In the dilute limit, the interaction
energy density is given by 
\begin{equation}
\mathcal{E}=\frac{2\pi\hbar^{2}a_{{\rm ad}}}{m_{{\rm ad}}}n_{{\rm b}}n_{{\rm d}},\label{Ead}
\end{equation}
where $n_{{\rm d}}$ is density of dimers, $n_{{\rm d}}=n_{\text{f}}/2$,
and $m_{{\rm ad}}$ is the reduced mass of the bosonic atom and dimer.

Equations (\ref{Ebf}) and (\ref{Ead}) are two different ways of representing
the same interaction energy. Therefore, by equating these two expressions,
we can obtain the atom-dimer scattering length from mean-field theory
as 
\begin{equation}
a_{{\rm ad}}^{0}=\frac{2m_{{\rm ad}}}{m_{{\rm bf}}}a_{{\rm bf}}.\label{amean}
\end{equation}
In this paper we will determine precisely the atom-dimer scattering
length from a three-body calculation. Its deviation from Eq. (\ref{amean})
will be used as a calibration of the mean-field theory. Our calculation
focuses on the resonance regime of Fermi superfluid with $a_{\text{bf}}/a_{\text{f}}\ll1$
(where $a_{\text{f}}$ denotes the scattering length between two fermionic
components). Besides, we also retain ourself within the case of $a_{\text{bf}}>0$
to ensure the stability of this system) \cite{Cui}. To our surprise,
it is found that even for weak boson-fermion interaction where $a_{\text{bf}}/a_{{\rm f}}$
is only a few percent (for instance, for typical fermion density $n_{\text{f}}=1.33\times10^{13}{\rm cm}^{-3}$,
$a_{\text{bf}}=40.8a_{0}$, and when $1/(k_{\text{F}}a_{{\rm f}})=1$,
$a_{\text{bf}}/a_{{\rm f}}=0.01$), the difference between outcomes
of three-body calculation and mean-field approach can be as large
as $10\%$ - $30\%$, depending on the mass ratio of boson to fermion.
Our results give a correction to mean-field interaction energy on
the BEC side of the fermionic Feshbach resonance, and can be verified
by a frequency shift of collective dipole oscillation, as done in a recent
experiment \cite{exp}.

\section{Atom-Dimer Scattering Length}

\subsection{STM-equation approach}

The atom-dimer scattering length can be obtained by solving the Schrödinger
equation for the system composed of a bosonic atom and two fermionic
atoms with different spins. A similar method has been applied to a two-component
Fermi system to obtain the fermion-dimer scattering length 1.2$a_{{\rm s}}$
\cite{petrov03,petrov05}. In our case, we label the two fermions
as 1 and 2, and the boson as atom 3. The Hamiltonian for this three-body
system is then given by $H=T+V_{12}+V_{23}+V_{31}\equiv T+V$, where
$T$ is the kinetic energy and $V_{ij}$ is the interaction potential
between atoms $i$ and $j$. We model the interaction $V_{ij}$ by the
Huang-Yang pseudopotential \cite{huangyang}
\begin{equation}
V_{ij}=\frac{2\pi\hbar^{2}a_{ij}}{m_{ij}}\delta({\bf r}_{ij})\frac{\partial}{\partial r_{ij}}(r_{ij}\cdot),\label{vij}
\end{equation}
with $a_{ij}$ the scattering length between the atoms $i$ and $j$
(i.e., $a_{12}=a_{{\rm f}}$ and $a_{31}=a_{23}=a_{{\rm bf}}$), and
${\bf r}_{ij}$ and $m_{ij}$ the relative coordinate and reduced
mass of these two atoms, respectively.

When $a_{{\rm f}}>0$, two fermionic atoms can form a dimer with binding
energy $\hbar^{2}/(2m_{12}a_{{\rm f}}^{2})$ and bound-state wave
function $|\phi_{b}\rangle_{12}$. The atom-dimer scattering length
$a_{{\rm ad}}$ is defined as \cite{sc} 
\begin{equation}
a_{{\rm ad}}=4\pi^{2}\hbar m_{{\rm ad}}\langle\Psi_{{\rm in}}|(V_{23}+V_{31})|\Psi_{+}\rangle.\label{aad}
\end{equation}
Here, $|\Psi_{{\rm in}}\rangle$ is the incident state of the atom-dimer
scattering process which can be expressed as $|\Psi_{{\rm in}}\rangle=|\phi_{b}\rangle_{12}|0\rangle_{3-12}$,
with $|0\rangle_{3-12}$ the eigenstate of the relative momentum between
atom 3 and the center of mass of atoms 1 and 2, with eigenvalue zero.
In Eq. (\ref{aad}), $|\Psi_{+}\rangle$ is the three-body scattering
state related to $|\Psi_{{\rm in}}\rangle$ via the Lippmann-Schwinger
equation \cite{sc}:
\begin{equation}
|\Psi_{+}\rangle=\lim_{\varepsilon\rightarrow0^{+}}\frac{i\varepsilon}{-\frac{\hbar^{2}}{2m_{{\rm 12}}a_{{\rm f}}^{2}}+i\varepsilon-H}|\Psi_{{\rm in}}\rangle.\label{psip}
\end{equation}

With a straightforward calculation employing Eqs. (\ref{aad}) and
(\ref{psip}) we obtain the Skorniakov-Ter-Martirosian (STM) equations
\cite{stm} for our system (Appendix A). In the natural unit $\hbar=m_{{\rm f}}=1$
with $m_{{\rm f}}$ the mass of a fermionic atom, these equations
are \begin{widetext}

\begin{eqnarray}
 &  & \frac{\frac{M''}{M}a(K,\varepsilon)}{\frac{1}{a_{{\rm f}}}+\sqrt{\frac{M''}{M}K^{2}+\frac{1}{a_{{\rm f}}^{2}}-i\varepsilon}}+\frac{2}{\pi^{2}}\int_{0}^{\Lambda e^{i\eta}}dK'\frac{K'}{K}{\rm ln}\left[\frac{\frac{M'}{2M}K^{2}+KK'+\gamma_{K'}}{\frac{M'}{2M}K^{2}-KK'+\gamma_{K'}}\right]\zeta(K',\varepsilon)=0;\label{stm1-1}\\
 &  & \left[\frac{1}{a_{{\rm bf}}}-\frac{\sqrt{2MM'}}{M'}\sqrt{\frac{M''K^{2}}{2M'}+\frac{1}{a_{{\rm f}}^{2}}-i\varepsilon}\right]\zeta(K,\varepsilon)-\frac{2M'}{M}\int_{0}^{\Lambda e^{i\eta}}dK'\frac{K'}{K[K'^{2}-i(\frac{4M}{M''})\varepsilon]}{\rm ln}\left[\frac{\frac{M'K'^{2}}{2M}+KK'+\gamma_{K}}{\frac{M'K'^{2}}{2M}-KK'+\gamma_{K}}\right]a(K',\varepsilon)\nonumber \\
 &  & +\frac{M'}{2\pi}\int_{0}^{\Lambda e^{i\eta}}dK'\frac{K'}{K}{\rm ln}\left[\frac{\frac{M'}{2M}(K^{2}+K'^{2})+\frac{1}{M}KK'+\frac{1}{a_{{\rm f}}^{2}}-i\varepsilon}{\frac{M'}{2M}(K^{2}+K'^{2})-\frac{1}{M}KK'+\frac{1}{a_{{\rm f}}^{2}}-i\varepsilon}\right]\zeta(K',\varepsilon)=\frac{2\pi M'}{M}\left[\frac{i\varepsilon}{\gamma_{K}(\gamma_{K}+i\varepsilon)}-\frac{1}{\gamma_{K}}\right],\label{stm2-1}
\end{eqnarray}
\end{widetext} where $M=m_{{\rm b}}/m_{{\rm f}}$ with $m_{{\rm b}}$
the mass of a bosonic atom, $M'=M+1,$ $M''=M+2$ and $\gamma_{K}=K^{2}+a_{{\rm f}}^{-2}-i\varepsilon$.
As shown in Appendix A, after solving Eqs. (\ref{stm1-1}) and (\ref{stm2-1})
we can obtain the atom-dimer scattering length $a_{{\rm ad}}$ via
the relation $a_{{\rm ad}}=\lim_{\varepsilon\rightarrow0^{+}}a(0,\varepsilon)$.
The atom-dimer scattering length has also been studied in other systems
\cite{ad-petrov,ad-mora,ad-levinsen09,ad-Nakajima,ad-Lompe,ad-Hammer,ad-Alzetto10,ad-Braaten,ad-Levinsen,ad-Alzetto12,ad-Ngampruetikorn,newaad}.

\begin{figure}
\includegraphics[width=5.0cm]{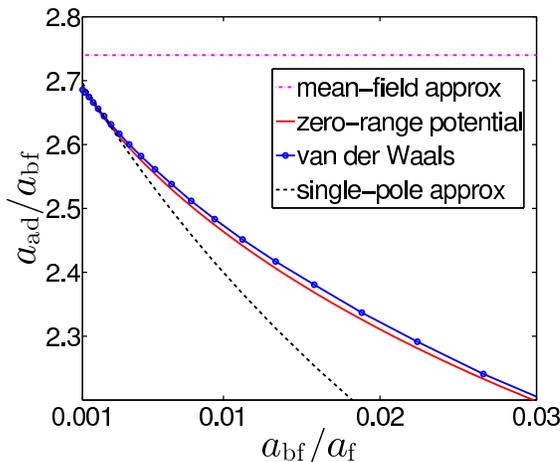}
 \caption{(Color online) The atom-dimer scattering length $a_{{\rm ad}}/a_{{\rm f}}$ 
as a function of $a_{\text{bf}}/a_{{\rm f}}$ given by three-body
calculation with mean-field approximation (purple dash-dotted line),
zero-range pseudo potential (red solid line), single-pole approximation
(black dashed line), and three-body calculation with a separable potential
suggested in Ref. \cite{pascal sepe} to incorporate the van der Waals
effect (blue circle with solid line). Here we illustrate the results for $^{6}$Li-$^{7}$Li mixture with $a_{{\rm bf}}=40.8a_{0}$, $\Lambda=4/a_{{\rm bf}}$, $\eta=0$,
$R_{{\rm vdW}}=31.26a_{0}$ for $^{6}$Li-$^{6}$Li interaction, and
$R_{{\rm vdW}}=32.49a_{0}$ for $^{6}$Li-$^{7}$Li interaction \cite{Rvdw-yan,Rvdw-chin,Rvdw-paul}.
\label{ad}}
\end{figure}

We note that in Eqs. (\ref{stm1-1}) and (\ref{stm2-1}), a high-energy
cutoff $\Lambda{\rm e}^{i\eta}$ is required in order to regularize
integrations. The exact value of $\Lambda$ and $\eta$ is determined
by the detail of van der Waals interaction potential between two atoms.
It is now known that $\Lambda$ is of the order of the van der Waals
length \cite{three-body-hutson,three-body-cheng,three-body-jia,three-body-zwerger,three-body-yujun,pascal sepe,pascal early},
and $\eta$ is usually very small. Since for weak boson-fermion interaction,
$a_{\text{bf}}$ is also of the order of the van der Waals length,
we have varied $\Lambda$ between $2/a_{{\rm bf}}$ and $6/a_{{\rm bf}}$
and varied $\eta$ between 0 and 0.08.

It is found that the final
result in the regime we are interested in is very insensitive to $\Lambda a_{\text{bf}}$
and $\eta$.
We also note that in the zero-range model, a boson can form
a two-body bound state with one of the fermions with binding energy
$1/(2m_{{\rm bf}}a_{{\rm bf}}^{2})$. It is then possible for a dimer
composed of two fermions to break up after scattering with a boson,
and one fermion paired with this boson to form such a boson-fermion
bound state. This inelastic atom-dimer scattering process will also
give rise to an imaginary part for $a_{\text{ad}}$. However, within
the regime we are interested in, the boson-fermion bound state is
deeply bounded with very small overlap with the incident state. As
a consequence, the inelastic scattering cross-section is extremely
small and the imaginary part of $a_{\text{ad}}$ remains negligible.

The results of $a_{\text{ad}}$ for a $^{6}$Li-$^{7}$Li mixture is
shown by the red solid line in Fig. \ref{ad}. Here we fix $a_{\text{bf}}$
and calculate $a_{\text{ad}}/a_{\text{bf}}$ with varying $a_{\text{bf}}/a_{\text{f}}$.
Comparing with the mean-field result shown by the horizontal dash-dotted
line, one can see that the two approaches agree with each other only
when $a_{\text{bf}}/a_{\text{f}}\rightarrow0$. For finite and positive
$a_{\text{bf}}/a_{\text{f}}$, the three-body result is always below
the mean-field expectation. Specifically, the deviation is already
about 10\% for $a_{\text{bf}}/a_{\text{f}}\approx0.01$, and
keeps increasing with $a_{\text{bf}}/a_{\text{f}}$ until $a_{\text{bf}}$
becomes comparable to $a_{\text{f}}$ where Efimov physics starts
to set in.

\begin{figure}
\includegraphics[width=5.0cm]{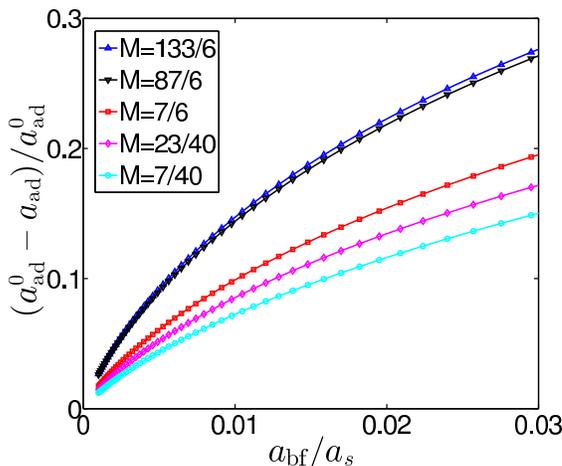} \caption{(Color online) The relative deviation $(a_{{\rm ad}}^{0}-a_{{\rm ad}})/a_{{\rm ad}}^{0}$
for different mass ratios ($M\equiv m_{{\rm b}}/m_{{\rm f}}$). Cases
shown are, from top to bottom, $^{133}$Cs-$^{6}$Li ($133/6$, blue
line with triangle), $^{87}$Rb-$^{6}$Li ($87/6$, black line with
inverted triangle), $^{7}$Li-$^{6}$Li ($7/6$, red line with square),
$^{23}$Na-$^{40}$K ($23/40$, purple line with diamond), and $^{7}$Li-$^{40}$K
($7/40$, cyan line with circle). \label{mass}}
\end{figure}

In Fig. \ref{mass} we investigate this deviation for other possible
realizations of Bose-Fermi mixtures, including $^{6}$Li-$^{87}$Rb,
$^{6}$Li-$^{133}$Cs, $^{40}$K-$^{23}$Na, and $^{40}$K-$^{7}$Li.
Here, we plot the relative derivation of the three-body result from
the mean-field value as $(a_{\text{ad}}^{0}-a_{\text{ad}})/a_{\text{ad}}^{0}$.
We find, on one hand, they all have qualitatively the same behaviors,
and on the other hand, the deviation increases with enhanced boson-fermion
mass ratio.

In the following, we would like to further understand two questions:
first, what is the major physical process that causes such a significant
deviation; second, since $a_{\text{bf}}$ is comparable to van der
Waals length, whether a van der Waals potential will change this result
obtained from zero-range models. The first question is to understand
the underlying physics better, while the second question is crucial
when comparing with experiments.

\subsection{Born and single-pole approximations}

To understand more about the difference between $a_{\text{ad}}$ and
$a_{\text{ad}}^{0}$, we can rewrite Eq. (\ref{psip}) as
\begin{equation}
|\Psi_{+}\rangle=|\Psi_{{\rm in}}\rangle+G_{3}(V_{23}+V_{31})|\Psi_{+}\rangle,\label{eq:-24}
\end{equation}
where $G_{3}=[-\hbar^{2}/(2m_{12}a_{{\rm f}}^{2})+i0^{+}-(T+V_{12})]$
is the Green's operator for a free boson and two interacting fermions,
and expand Eq. (\ref{aad}) as 
\begin{align}
 & a_{{\rm ad}}=4\pi^{2}\hbar m_{{\rm ad}}\langle\Psi_{{\rm in}}|(V_{23}+V_{31})|\Psi_{{\rm in}}\rangle\nonumber \\
 & +4\pi^{2}\hbar m_{{\rm ad}}\langle\Psi_{{\rm in}}|(V_{23}+V_{31})G_{3}(V_{23}+V_{31})|\Psi_{{\rm in}}\rangle+\cdots\label{aad2}
\end{align}
If we take the first-order Born approximation by neglecting all higher-order terms in Eq. (\ref{aad2}), it is straightforward to show that
\begin{align}
a_{{\rm ad}} & =4\hbar\pi^{2}m_{{\rm ad}}\langle\Psi_{{\rm in}}|(V_{23}+V_{31})|\Psi_{{\rm in}}\rangle=\frac{2m_{{\rm ad}}}{m_{{\rm bf}}}a_{{\rm bf}}=a_{{\rm ad}}^{0}.\nonumber \\
\label{eq:-26}
\end{align}
This means that $a_{\text{ad}}^{0}$ deduced from many-body mean-field
treatment is equivalent to that obtained from a three-body calculation
with first-order Born approximation. Thus, the difference between
the exact $a_{{\rm ad}}$ and the mean-field $a_{\text{ad}}^{0}$
is due to processes beyond the first-order Born approximation.

The higher-order terms of Eq. (\ref{aad2}) correspond to the following
two types of atom-dimer scattering processes: (i) a dimer composed
of two fermions remains in the bound state, and undergoes repeated
collisions with bosons; (ii) an incoming dimer first breaks into two
fermions in the scattering state, and then they return to a bound
state after the second collision. While the processes of the second
type are difficult to incorporate, the ones of the first type can
be integrated using a ``single-pole'' approximation. Within this
approach, the three-body Green's function $G_{3}$ in Eq. (\ref{aad2})
is approximated as
\begin{equation}
G_{3}\approx\int d{\bf K}\frac{|{\bf K}\rangle_{3-12}\langle{\bf K}|\otimes|\phi_{b}\rangle_{12}\langle\phi_{b}|}{i0^{+}-\frac{K^{2}}{2m_{{\rm ad}}}},\label{eq:-25}
\end{equation}
where $|{\bf K}\rangle_{3-12}$ is the eigenstate of the relative
momentum between atom 3 and the center of mass of atoms 1 and 2. As
a result, the wave function for relative motion between atoms $1$
and $2$ is forced in the bound state. With this approximation, process
(i) is fully taken into account, while process (ii) is neglected.

As shown in Appendix B, under the single-pole approximation we can
obtain the atom-dimer scattering length $a_{{\rm ad}}$ via the relation
($\hbar=m_{{\rm f}}=1$) $a_{{\rm ad}}=\lim_{\varepsilon\rightarrow0^{+}}\left(\frac{8M\pi^{2}}{M+2}\right)T(0,\varepsilon)$.
Here the function $T(K,\varepsilon)$ satisfies the integral equation
\begin{eqnarray}
 &  & T(K,\varepsilon)+\frac{8(M+1)a_{{\rm bf}}}{(M+2)\pi^{2}a_{{\rm f}}}\int dK'{\cal F}(K,K',\varepsilon)T(K',\varepsilon)\nonumber \\
 & = & \frac{(M+1)a_{{\rm bf}}}{M\pi^{3}a_{{\rm f}}}\int_{0}^{\infty}\frac{k'dk'}{K\gamma_{k'}}\log\left[\frac{4+a_{{\rm f}}^{2}(K+2k')^{2}-4i\varepsilon}{4+a_{{\rm f}}^{2}(K-2k')^{2}-4i\varepsilon}\right],\nonumber \\
\label{spa}
\end{eqnarray}
where the function ${\cal F}(K,K',\varepsilon)$ is defined as 
\begin{eqnarray}
{\cal F}(K,K',\varepsilon) & = & \int_{-1}^{1}du\int_{0}^{\infty}dk'\frac{k'}{\zeta(K,K',u)\gamma_{K'}}\times\nonumber \\
 &  & \log\left[\frac{4+a_{{\rm f}}^{2}[\zeta(K,K',u)+2k']^{2}-4i\varepsilon}{4+a_{{\rm f}}^{2}[\zeta(K,K',u)-2k']^{2}-4i\varepsilon}\right],\nonumber \\
\label{eq:-27}
\end{eqnarray}
with $\zeta(K,K',u)=\sqrt{K^{2}-2uKK'+K'^{2}}$ and $\gamma_{k'}=k'^{2}+a_{{\rm f}}^{-2}-i\varepsilon$.

The result of $a_{\text{ad}}$ from this single-pole approximation
is shown as the black dashed line in Fig. \ref{ad} for the $^{6}$Li-$^{7}$Li
mixture, which has the same qualitative behavior as the exact three-body
calculation and also gives a large deviation from the mean-field value
$a_{\text{ad}}^{0}$. This suggests that processes of type (i) are
important processes which significantly reduce $a_{\text{ad}}$
from $a_{\text{ad}}^{0}$. This also shows that in a more accurate
many-body theory, it is necessary to include the ladder diagram to
describe repeated scattering between fermion pairs and bosons. It
is reminiscent of the ladder diagram between fermion pairs that reduces
dimer-dimer scattering length significantly below its mean-field value
on the BEC side of the Feshbach resonance \cite{pair_flucuation}.

\subsection{Effect of van der Waals potential}

To investigate the effect of van der Waals potential, we implement
the separable potential proposed in Ref. \cite{pascal sepe}. In this
method, the interaction potential between atoms $i$ and $j$ is modeled
as
\begin{equation}
V_{ij}=\xi|\chi\rangle_{ij}\langle\chi|,\label{eq:-1}
\end{equation}
with $|\chi\rangle_{ij}$ a state for the relative motion of these
two atoms. The potential $V_{ij}$ is designed to reproduce not only
the two-atom scattering length $a_{ij}$ but also the $s$-wave zero-energy
scattering wave function $u(r)/r$ in a van der Waals potential $-C_{6}/r^{6}$.
To meet this requirement, the state $|\chi\rangle_{ij}$ and the parameter
$\xi$ should be chosen as $_{ij}\langle{\bf q}|\chi\rangle_{ij}=(2\pi)^{-3/2}\{1-q\int_{0}^{\infty}dr[1-\frac{r}{a_{ij}}-u(r)]\sin(qr)\}$
and $\xi=4\pi[\frac{1}{a_{ij}}-\frac{2}{\pi}\int_{0}^{\infty}dq|_{ij}\langle{\bf q}|\chi\rangle_{ij}|^{2}]^{-1}$,
respectively \cite{pascal sepe}, where $|{\bf q}\rangle_{ij}$ is
the eigen-state of the relative momentum with eigen-value ${\bf q}$.
This separable potential naturally includes the length scale of the
van der Waals length $R_{{\rm vdW}}=1/2(mC_{6}/\hbar^{2})^{1/4}$
through wave function $u({\bf r})$. Besides, the three-body calculation
with this potential does not require extra three-body parameters.
Indeed, Ref. \cite{pascal sepe} used this model to show the three-body
parameter depends on the van der Waals length universally. 

Here we use this separable potential to calculate the atom-dimer
scattering length $a_{{\rm ad}}$ defined in Eq. (\ref{aad}). Following
the same strategy as in Appendix A, we straightforwardly derive integral
equations which are quite similar to STM equations, and obtain $a_{{\rm ad}}$
via numerically solving these equations. Here we have used the facts
that $R_{{\rm vdW}}=31.26a_{0}$ for $^{6}$Li-$^{6}$Li interaction,
and $R_{{\rm vdW}}=32.49a_{0}$ for $^{6}$Li-$^{7}$Li interaction
\cite{Rvdw-yan,Rvdw-chin,Rvdw-paul}. The result of $a_{\text{ad}}$
from this separable potential is also shown in Fig. \ref{ad}. We
find that in the regime of our interest, the correction is visible
but not significant.

\section{Experimental Predictions}

With three-body calculation of $a_{\text{ad}}$, we can provide a
correction to the mean-field interaction energy between Bose and Fermi
superfluids on the BEC side of the fermionic Feshbach resonance. This
interaction energy has been extracted from a measurement of collective
mode frequency shift in the underlying system \cite{exp}, where the
condensate of bosonic atoms with smaller particle number is embedded
in a larger cloud of Fermi superfluid. So the effective trapping potential
experienced by bosons should include contributions from both the harmonic
trap $V(\text{{\bf r}})=(1/2)m\omega_{{\rm b}}^{2}{\bf r}^{2}$ and
the interaction energy between Bose and Fermi superfluids, leading
to $V_{\text{eff}}({\bf r})=V({\bf r})+2\pi\hbar^{2}a_{{\rm ad}}n_{{\rm d}}(\text{{\bf r}})/m_{{\rm ad}}$
on the BEC side of the Feshbach resonance. Here, since the number
of fermions is much larger than that of bosons, we can safely assume
that the fermion density distribution (i.e. dimer density distribution
$n_{{\rm d}}$) will not be affected by bosons and is solely determined
by the equation of state of Fermi superfluid. Such an equation of
state has been obtained quite accurately in previous experiment \cite{eos}.
With the local density approximation, we have $n_{{\rm d}}(\text{{\bf r}})=n_{{\rm d}}[\mu_{{\rm f}}-V(\text{{\bf r}})]$,
where we have used the fact that $^{6}$Li and $^{7}$Li experience
almost identical trapping potentials. Thus, for the dipole oscillation
of bosons, its frequency $\tilde{\omega}_{{\rm b}}$ will be shifted
away from $\omega_{{\rm b}}$, and the shift $\delta\omega_{{\rm b}}=\omega_{{\rm b}}-\tilde{\omega}_{{\rm b}}$
is given by \cite{exp} 
\begin{equation}
\frac{\delta\omega_{{\rm b}}}{\omega_{{\rm b}}}=\frac{\pi\hbar^{2}a_{{\rm ad}}}{m_{{\rm ad}}}\left(\frac{dn_{{\rm d}}}{d\mu_{{\rm f}}}\right)_{\text{{\bf r}}={\rm 0}}.\label{domega}
\end{equation}

\begin{figure}
\includegraphics[width=5.0cm]{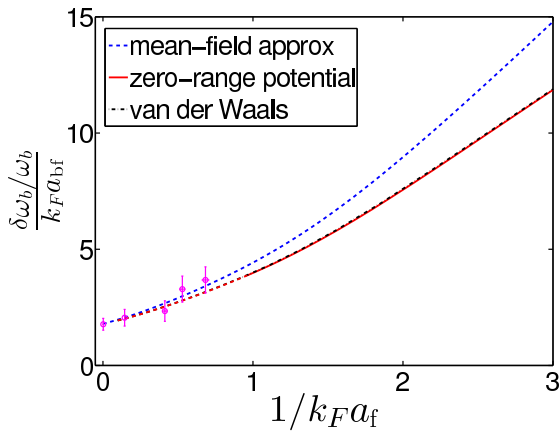} \caption{(Color online) Frequency shift of dipole oscillation $\delta\omega_{{\rm b}}$
for bosonic $^{7}$Li atoms as a function of $1/(k_{\text{F}}a_{\text{f}})$.
Dots with error bar are experimental data \cite{exp}. Blue dashed line
is predication from mean-field theory. Red solid line is predication
based on $a_{\text{ad}}$ calculated with zero-range model, and black
dash-dotted line is predication based on $a_{\text{ad}}$ from separable
potential including the van der Waals effect. Here we have chosen $k_{F}=4.6\times10^{6}m^{-1}$.
Other parameters take the same values as in Fig. 1. \label{exp}}
\end{figure}

In Ref. \cite{exp}, this collective frequency has been measured in
the unitary regime (with $1/(k_{\text{F}}|a_{\text{f}}|)<1$) and
fitted with a mean-field theory by replacing $a_{\text{ad}}$ with
$a_{\text{ad}}^{0}$ in Eq. (\ref{domega}). The experimental data
points and the mean-field fitting are shown in Fig. \ref{exp}. However,
in this regime, the size of dimers is even larger than the interparticle
distance and the system can not be viewed as a boson-dimer mixture.
Our expression of Eq. (\ref{domega}) should be applied to the regime
with $1/(k_{\text{F}}a_{\text{f}})>1$, where the results for $\delta\omega_{{\rm b}}/\omega_{{\rm b}}$
with both zero-range and separable potential are shown in Fig. \ref{exp}
and compared with mean-field predication. Future experiments can perform
more accurate measurements of frequency shift in the BEC regime; the
difference between our result and the mean-field result can be distinguished.

\section{Summary}

In summary, our work studies the interaction energy between Bose and
Fermi superfluids on the BEC side of the fermionic Feshbach resonance.
Our result provides a striking example of how mean-field theory can
be qualitatively inaccurate even for quite weak interaction strength,
and suggests a route to improve many-body theory in this system. This
result can be easily verified in the current experimental setup. 
\begin{acknowledgments}
We wish to thank Pascal Naidon, Ran Qi, Christophe Salomon, Frederic
Chevy, Yujun Wang, and Xiaoling Cui for helpful discussions. This
work is supported by the NSFC under Grants No. 1127400 (W.Z.), No. 11174176(H.Z.) and 
No. 11222430 (P.Z.); the NKBRP under Grants No. 2013CB922000 (W.Z.), No. 2011CB921500
(H.Z.), and No. 2012CB922104 (P.Z.); the Program of State Key Laboratory of Quantum
Optics and Quantum Optics Devices under Grant No. KF201404 (W.Z.); and the Tsinghua University
Initiative Scientific Research Program (H.Z.). 
\end{acknowledgments}
\appendix
\addcontentsline{toc}{section}{Appendices}\markboth{APPENDICES}{}

\section{STM Equations}

In this Appendix we show our STM-equation approach for the calculation
of atom-dimer scattering length $a_{{\rm ad}}$. We will first introduce
a function $a({\bf K},\varepsilon)$, and then prove that such a function
satisfies the relation 
\begin{equation}
\lim_{\varepsilon\rightarrow0^{+}}a(0,\varepsilon)=a_{{\rm ad}}.\label{rel}
\end{equation}
Namely, one can obtain $a_{{\rm ad}}$ directly from $a({\bf K},\varepsilon)$.
Then we will derive the equation for $a({\bf K},\varepsilon)$,
i.e., Eqs. (\ref{stm1-1}) and (\ref{stm2-1}) in our main text. They
are the STM equations of our system.

\subsection{Function $a({\bf K},\varepsilon)$}

As shown in the main text, our system includes two fermionic atoms
(labeled 1 and 2) with the same mass and different spins, and one bosonic
atom (labeled 3). In the following we use $|\cdot\rangle$ to denote
the total quantum state of the relative motion of these three atoms,
$|\cdot\rangle_{ij}$ for the state of the relative motion between
atoms $i$ and $j$, and $|\cdot\rangle_{i-jk}$ for the state of
the relative motion between the atom $i$ and the center of mass of
atoms $j$ and $k$. We further denote the mass of the fermionic atom
and bosonic atom as $m_{{\rm f}}$ and $m_{{\rm b}}$, respectively.
In our calculation we use the natural unit $\hbar=m_{{\rm f}}=1$.
We model the two-body interaction with Huang-Yang pseudo potential,
i.e., the interaction operator $V_{ij}$ for the atoms $i$ and $j$
satisfies 
\begin{equation}
_{ij}\langle{\bf r}|V_{ij}|\Psi\rangle=\frac{2\pi a_{ij}}{m_{ij}}\delta({\bf r})\left[\frac{\partial}{\partial|{\bf r}|}\left(|{\bf r}|\times{}_{ij}\langle{\bf r}|\Psi\rangle\right)\right],\label{potential-2}
\end{equation}
where $|{\bf r}\rangle_{ij}$ is the eigenstate of the relative position
between atoms $i$ and $j$ with eigenvalue ${\bf r}$ , and $m_{ij}$
and $a_{ij}$ are the reduced mass and scattering length of these
two atoms, respectively. As shown in the main text, here we assume
$a_{12}=a_{{\rm f}}$ and $a_{31}=a_{23}=a_{{\rm bf}}$.

To introduce the function $a({\bf K},\varepsilon)$, we first define
state $|\psi_{3}(\varepsilon)\rangle_{3-12}$ via the relation 
\begin{equation}
_{12}\langle{\bf r}|V_{12}|\Psi_{+}(\varepsilon)\rangle=\delta({\bf r})|\psi_{3}(\varepsilon)\rangle_{3-12},\label{psiijk}
\end{equation}
where $|\Psi_{+}(\varepsilon)\rangle$ is defined as 
\begin{equation}
|\Psi_{+}(\varepsilon)\rangle=\frac{i\varepsilon}{-{a_{{\rm f}}^{-2}}+i\varepsilon-(T+V_{12}+V_{23}+V_{31})}|\Psi_{{\rm in}}\rangle\label{psipeps}
\end{equation}
with $T$ the kinetic energy of the relative motion of these three
atoms and $|\Psi_{{\rm in}}\rangle$ the incident state of the atom-dimer
scattering process, as defined in the main text. We further define
function $\eta({\bf K},\varepsilon)$ as 
\begin{equation}
\eta({\bf K},\varepsilon)={}_{3-12}\langle{\bf K}|\psi_{3}(\varepsilon)\rangle_{3-12}.\label{eq:-10}
\end{equation}

With the aid of Eqs. (\ref{psiijk}), (\ref{psipeps}), and (\ref{eq:-10}),
we can define the function $a({\bf K},\varepsilon)$ via the equation
\begin{equation}
\eta({\bf K},\varepsilon)=-\frac{\sqrt{2}}{\pi^{3/2}a_{{\rm f}}^{1/2}}\left[2\pi^{2}\delta({\bf K})+\frac{a({\bf K},\varepsilon)}{\left(\frac{4M}{M+2}\right)i\varepsilon-K^{2}}\right],\label{eq:-3}
\end{equation}
where $M=m_{{\rm b}}/m_{{\rm f}}$ and $|{\bf K}\rangle_{k-ij}$ is
the eigenstate of the relative momentum between atom $k$ and the
center of mass of atoms $i$ and $j$, with eigenvalue ${\bf K}$.

\subsection{Relation Between $a({\bf K},\varepsilon)$ and $a_{{\rm ad}}$}

Now we prove the relation of Eq. (\ref{rel}) which links the function
$a({\bf K},\varepsilon)$ defined in Eq. (\ref{eq:-3}) with the atom-dimer
scattering length $a_{{\rm ad}}$. To this end we rewrite Eq. (\ref{psipeps})
as 
\begin{equation}
|\Psi_{+}(\varepsilon)\rangle=|\Psi_{{\rm in}}\rangle+G_{3}(\varepsilon)(V_{23}+V_{31})|\Psi_{+}(\varepsilon)\rangle,\label{eq:-2}
\end{equation}
where $G_{3}(\varepsilon)=[-a_{{\rm f}}^{-2}+i\varepsilon-(T+V_{12})]^{-1}$
is the Green's operator for the free bosonic atom together with the
two interacting fermionic atoms taking the form 
\begin{eqnarray}
G_{3}(\varepsilon) & = & \int d{\bf K}\frac{|{\bf K}\rangle_{3-12}\langle{\bf K}|\otimes|\phi_{b}\rangle_{12}\langle\phi_{b}|}{i\varepsilon-\left(\frac{M+2}{4M}\right){\bf K}^{2}}\nonumber \\
 &  & +\int d{\bf K}d{\bf k}\frac{|{\bf K}\rangle_{3-12}\langle{\bf K}|\otimes|{\bf k}+\rangle_{12}\langle{\bf k}+|}{-a_{{\rm f}}^{-2}+i\varepsilon-\left(\frac{M+2}{4M}\right){\bf K}^{2}-{\bf k}^{2}}.\nonumber \\
\label{g3}
\end{eqnarray}
Here $|{\bf k}+\rangle_{12}$ is the scattering state of the relative
motion between atoms $1$ and $2$ with incident momentum ${\bf k}$,
and $|\phi_{b}\rangle_{12}$ is the bound state of these two atoms,
as defined in our main text. By writing down Eq. (\ref{g3}), we have
used the fact that $T={\bf p}_{12}^{2}+\left(\frac{M+2}{4M}\right){\bf P}_{3-12}^{2},$
where ${\bf p}_{12}$ is the relative momentum operator between atoms
$1$ and $2$, and ${\bf P}_{3-12}$ is the relative momentum operator
of atom $3$ and the center of mass of atoms $1$ and $2$. We have
also used the eigen equations 
\begin{eqnarray}
\left({\bf p}_{12}^{2}+V_{12}\right)|\phi_{b}\rangle_{12} & = & -\frac{1}{a_{{\rm f}}^{2}}|\phi_{b}\rangle,\\
\left({\bf p}_{12}^{2}+V_{12}\right)|{\bf k}+\rangle_{12} & = & {\bf k}^{2}|{\bf k}+\rangle
\end{eqnarray}
satisfied by $|{\bf k}+\rangle_{12}$ and $|\phi_{b}\rangle_{12}$.

Now we are at the stage to calculate $\eta({\bf K},\varepsilon)$.
According to Eqs. (\ref{potential-2}) and (\ref{psiijk}), we have
\begin{equation}
\eta({\bf K},\varepsilon)=4\pi a_{{\rm f}}\left\{ \left.\frac{\partial}{\partial|{\bf r}|}\left[|{\bf r}|\times{}_{12}\langle{\bf r}|\left(_{3-12}\langle{\bf K}|\Psi_{+}(\varepsilon)\rangle\right)\right]\right|_{{\bf r}=0}\right\} .\label{rel1}
\end{equation}
Substituting Eqs. (\ref{eq:-2}) and (\ref{g3}) into Eq. (\ref{rel1}),
and using the results in two-body problems 
\begin{eqnarray}
_{12}\langle{\bf r}|\phi_{b}\rangle_{12} & = & \frac{1}{\sqrt{2a_{{\rm f}}\pi}}\frac{e^{-|{\bf r}|/a_{{\rm f}}}}{|{\bf r}|},\label{eq:}\\
_{12}\langle{\bf r}|{\bf k}+\rangle_{12} & = & \frac{1}{(2\pi)^{3/2}}\left[e^{i{\bf k}\cdot{\bf r}}+\left(\frac{-1}{i|{\bf k}|+\frac{1}{a_{{\rm f}}}}\right)\frac{e^{i|{\bf k}||{\bf r}|}}{|{\bf r}|}\right],\nonumber \\
\end{eqnarray}
as well as the fact that $|\Psi_{{\rm in}}\rangle=|\phi_{b}\rangle_{12}|0\rangle_{3-12}$,
with $|0\rangle_{3-12}$ the eigenstate of ${\bf P}_{3-12}$ with
eigenvalue zero, we obtain 
\begin{eqnarray}
 &  & \eta({\bf K},\varepsilon)=\nonumber \\
 &  & -\frac{2\sqrt{2\pi}}{\sqrt{a_{{\rm f}}}}\delta({\bf K})-\frac{2\sqrt{2\pi}}{\sqrt{a_{{\rm f}}}}\frac{\ _{12}\langle\phi_{b}|{}_{3-12}\langle{\bf K}|(V_{23}+V_{31})|\Psi_{+}(\varepsilon)\rangle}{i\varepsilon-\left(\frac{M+2}{4M}\right){\bf K}^{2}}\nonumber \\
 &  & -\int d{\bf k}\frac{ia_{{\rm f}}\sqrt{2}\times{}_{12}\langle{\bf k}+|{}_{3-12}\langle{\bf K}|(V_{23}+V_{31})|\Psi_{+}(\varepsilon)\rangle}{\sqrt{\pi}\left(-i+a_{{\rm f}}|{\bf k}|\right)\left[-\frac{1}{a_{{\rm f}}^{2}}+i\varepsilon-\left(\frac{M+2}{4M}\right){\bf K}^{2}-{\bf k}^{2}\right]}.\nonumber \\
\label{eq:-4}
\end{eqnarray}
Comparing Eq. (\ref{eq:-4}) with the definition of $a({\bf K},\varepsilon)$
Eq. (\ref{eq:-3}), we find that $a({\bf K},\varepsilon)$ can be
re-expressed as 
\begin{eqnarray}
a({\bf K},\varepsilon) & = & \left(\frac{8M\pi^{2}}{M+2}\right)\left[_{12}\langle\phi_{b}|{}_{3-12}\langle{\bf K}|(V_{23}+V_{31})|\Psi_{+}(\varepsilon)\rangle\right]\nonumber \\
 &  & +\left[\left(\frac{4M}{M+2}\right)i\varepsilon-{\bf K}^{2}\right]g({\bf K},\varepsilon)\label{eq:-6}
\end{eqnarray}
where 
\begin{eqnarray}
 &  & g({\bf K},\varepsilon)\nonumber \\
 & = & \int d{\bf k}\frac{ia_{{\rm f}}^{3/2}\pi\times{}_{12}\langle{\bf k}+|{}_{3-12}\langle{\bf K}|(V_{23}+V_{31})|\Psi_{+}(\varepsilon)\rangle}{\left(-i+a_{{\rm f}}|{\bf k}|\right)\left[a_{{\rm f}}^{-2}+i\varepsilon-\left(\frac{M+2}{4M}\right){\bf K}^{2}-{\bf k}^{2}\right]}.\nonumber \\
\end{eqnarray}
When ${\bf K}=0$, in the limit $\varepsilon\rightarrow0^{+}$ the
second term in the right-hand side of Eq. (\ref{eq:-6}) becomes zero.
Therefore, we have 
\begin{eqnarray}
 &  & \lim_{\varepsilon\rightarrow0^{+}}a(0,\varepsilon)\nonumber \\
 & = & \left(\frac{8M\pi^{2}}{M+2}\right)\lim_{\varepsilon\rightarrow0^{+}}\left[_{12}\langle\phi_{b}|{}_{3-12}\langle0|(V_{23}+V_{31})|\Psi_{+}(\varepsilon)\rangle\right]\nonumber \\
\label{eq:-7}
\end{eqnarray}

On the other hand, according to Eqs. (5) and (6) in our main text,
the atom-dimer scattering length $a_{{\rm ad}}$ is given by 
\begin{align}
a_{{\rm ad}} & =\lim_{\varepsilon\rightarrow0^{+}}\left(\frac{8M\pi^{2}}{M+2}\right){}_{12}\langle\phi_{b}|{}_{3-12}\langle0|(V_{23}+V_{31})|\Psi_{+}(\varepsilon)\rangle.\nonumber \\
\label{aad-1}
\end{align}
Here we have used the fact that in our natural unit with $m_{{\rm f}}=1$,
the atom-dimer reduced mass is $2M/(M+2)$. With Eq. (\ref{aad-1}),
we can rewrite Eq. (\ref{eq:-7}) as $\lim_{\varepsilon\rightarrow0^{+}}a(0,\varepsilon)=a_{{\rm ad}}.$
That is the relation in Eq. (\ref{rel}).

\subsection{STM Equations}

In the previous discussions, we show that the atom-dimer scattering length
$a_{{\rm ad}}$ can be obtained directly from the function $a({\bf K},\varepsilon)$.
Now we derive the equation for the function $a({\bf K},\varepsilon)$,
i.e., the STM equation. To this end, we first define states $|\psi_{1}(\varepsilon)\rangle_{1-23}$
and $|\psi_{2}(\varepsilon)\rangle_{2-31}$ via relations 
\begin{eqnarray}
_{23}\langle{\bf r}|V_{23}|\Psi_{+}(\varepsilon)\rangle & = & \delta({\bf r})|\psi_{1}(\varepsilon)\rangle_{1-23},\label{psiijk-1}\\
_{31}\langle{\bf r}|V_{31}|\Psi_{+}(\varepsilon)\rangle & = & \delta({\bf r})|\psi_{2}(\varepsilon)\rangle_{2-31}.\label{psiijk-1b}
\end{eqnarray}
In our system, since atoms $1$ and $2$ have the same mass and scattering
length with atom $3$, it is easy to prove that $_{1-23}\langle{\bf K}|\psi_{1}(\varepsilon)\rangle_{1-23}={}_{2-31}\langle{\bf K}|\psi_{2}(\varepsilon)\rangle_{2-31}$.
We further define the function $\zeta({\bf K},\varepsilon)$ as 
\begin{eqnarray}
\text{\ensuremath{\zeta}}({\bf K},\varepsilon) & = & -\left(2^{\frac{3}{2}}\pi^{\frac{5}{2}}\sqrt{a_{{\rm f}}}\right){}_{1-23}\langle{\bf K}|\psi_{1}(\varepsilon)\rangle_{1-23}\nonumber \\
 & = & -\left(2^{\frac{3}{2}}\pi^{\frac{5}{2}}\sqrt{a_{{\rm f}}}\right){}_{2-31}\langle{\bf K}|\psi_{2}(\varepsilon)\rangle_{2-31}.\nonumber \\
\label{eq:-8}
\end{eqnarray}
As we have outlined in the previous section, it is easy to prove that
\begin{eqnarray}
 &  & \zeta({\bf K},\varepsilon)\nonumber \\
 & = & -\text{(2\ensuremath{\pi})}^{\frac{7}{2}}a_{{\rm f}}^{\frac{1}{2}}a_{{\rm bf}}\left\{ \left.\frac{\partial}{\partial|{\bf r}|}\left[|{\bf r}|\times{}_{23}\langle{\bf r}|\left(_{1-23}\langle{\bf K}|\Psi_{+}(\varepsilon)\rangle\right)\right]\right|_{{\bf r}=0}\right\} \nonumber \\
\label{rel1-1}\\
 & = & -\text{(2\ensuremath{\pi})}^{\frac{7}{2}}a_{{\rm f}}^{\frac{1}{2}}a_{{\rm bf}}\left\{ \left.\frac{\partial}{\partial|{\bf r}|}\left[|{\bf r}|\times{}_{31}\langle{\bf r}|\left(_{2-31}\langle{\bf K}|\Psi_{+}(\varepsilon)\rangle\right)\right]\right|_{{\bf r}=0}\right\} \nonumber \\
\label{eq:-9}
\end{eqnarray}
by using Eqs. (\ref{potential-2}), (\ref{psiijk-1}), (\ref{psiijk-1b})
and (\ref{eq:-8}).

Now we rewirte Eq. (\ref{psipeps}) as 
\begin{eqnarray}
|\Psi_{+}(\varepsilon)\rangle=i\varepsilon G_{0}(\varepsilon)|\Psi_{{\rm in}}\rangle+G_{0}(\varepsilon)\left(V_{12}+V_{23}+V_{31}\right)|\Psi_{+}(\varepsilon)\rangle,\nonumber \\
\label{eqn:LP equation1-1}
\end{eqnarray}
where 
\begin{eqnarray}
G_{0}(\varepsilon) & = & \frac{1}{-{a_{{\rm f}}^{-2}}+i\varepsilon-T}\nonumber \\
 & = & \int d{\bf k}d{\bf K}\frac{|{\bf k}\rangle_{ij}\langle{\bf k}|\otimes|{\bf K}\rangle_{k-ij}\langle{\bf K}|}{-{a_{{\rm f}}^{-2}}+i\varepsilon-\left[|{\bf k}|^{2}+\left(\frac{M+2}{4M}\right)K^{2}\right]},\nonumber \\
\label{g0}
\end{eqnarray}
with $|{\bf k}\rangle_{ij}$ the eigenstate of ${\bf p}_{ij}$ with
eigenvalue ${\bf k}$ and $K=|{\bf K}|$. Here $(i,j,k)$ can take
values $(1,2,3)$, $(2,3,1)$, or $(3,1,2)$. Substituting Eq. (\ref{eqn:LP equation1-1})
into Eq. (\ref{rel1}), and using Eqs. (\ref{eq:-8}), (\ref{rel1-1})
and (\ref{eq:-9}), we obtain 
\begin{eqnarray}
 &  & \left(\frac{1}{a_{{\rm f}}}-\sqrt{\left(\frac{M+2}{4M}\right)K^{2}+\frac{1}{a_{{\rm f}}^{2}}-i\varepsilon}\right)\eta({\bf K},\varepsilon)\nonumber \\
 &  & -\int\frac{d{\bf K}'}{2^{\frac{3}{2}}\pi^{\frac{9}{2}}\sqrt{a_{{\rm f}}}}\frac{\zeta({\bf K}',\varepsilon)}{\left[\left(\frac{M+1}{2M}\right)K^{2}+K'^{2}+{\bf K}\cdot{\bf K}'+\frac{1}{a_{{\rm f}}^{2}}-i\varepsilon\right]}\nonumber \\
 &  & +\frac{2\sqrt{2\pi}}{\sqrt{a_{{\rm f}}}}\left(\frac{1}{a_{{\rm f}}}-\sqrt{\frac{1}{a_{{\rm f}}^{2}}-i\varepsilon}\right)\delta({\bf K})=0.\label{eq:-13}
\end{eqnarray}
Here we have used the relations

\begin{eqnarray}
 &  & _{12}\langle{\bf r}|\left(_{3-12}\langle{\bf K}|G_{0}(\varepsilon)V_{12}|\Psi_{+}(\varepsilon)\rangle\right)\nonumber \\
 & = & \frac{\eta({\bf K},\varepsilon)}{(2\pi)^{3}}\int d{\bf k}\frac{e^{i{\bf k}\cdot{\bf r}}}{-\frac{1}{a_{{\rm f}}^{2}}+i\varepsilon-\left[|{\bf k}|^{2}+\left(\frac{M+2}{4M}\right)K^{2}\right]}\nonumber \\
 & = & -\frac{\eta({\bf K},\varepsilon)}{(2\pi)^{3}}\int d{\bf k}\frac{e^{i{\bf k}\cdot{\bf r}}}{|{\bf k}|^{2}}+\frac{\eta({\bf K},\varepsilon)}{(2\pi)^{3}}\int d{\bf k}\left\{ \frac{e^{i{\bf k}\cdot{\bf r}}}{\Delta}+\frac{e^{i{\bf k}\cdot{\bf r}}}{|{\bf k}|^{2}}\right\} \nonumber \\
 & = & -\frac{\eta({\bf K},\varepsilon)}{32\pi^{4}|{\bf r}|}+\frac{\eta({\bf K},\varepsilon)}{(2\pi)^{3}}\int d{\bf k}\left\{ \frac{1}{\Delta}+\frac{1}{|{\bf k}|^{2}}\right\} +{\cal O}(|{\bf r}|),\nonumber \\
\label{eq:-11}
\end{eqnarray}
with $\Delta=-\frac{1}{a_{{\rm f}}^{2}}+i\varepsilon-\left[|{\bf k}|^{2}+\left(\frac{M+2}{4M}\right)K^{2}\right]$
and 
\begin{equation}
_{12}\langle{\bf k}|\phi_{b}\rangle_{12}=\frac{1}{\pi\sqrt{a_{{\rm f}}}}\frac{1}{\left(|{\bf k}|^{2}+\frac{1}{a_{{\rm f}}^{2}}\right)}.\label{eq:-12}
\end{equation}
Substituting Eq. (\ref{eq:-3}) into Eq. (\ref{eq:-13}), we obtain
\begin{eqnarray}
 &  & \int\frac{d{\bf K}'}{(2\pi)^{3}}\frac{\zeta({\bf K}',\varepsilon)}{\left[\left(\frac{M+1}{2M}\right)K^{2}+K'^{2}+{\bf K}\cdot{\bf K}'+\frac{1}{a_{{\rm f}}^{2}}-i\varepsilon\right]}\nonumber \\
 &  & +\frac{\left(\frac{M+2}{8M}\right)a({\bf K},\varepsilon)}{\frac{1}{a_{{\rm f}}}+\sqrt{\left(\frac{M+2}{4M}\right)K^{2}+\frac{1}{a_{{\rm f}}^{2}}-i\varepsilon}}=0.\label{eq:-14}
\end{eqnarray}
On the other hand, by substituting Eq. (\ref{eqn:LP equation1-1}) into
Eq. (\ref{rel1-1}) and using similar techniques as above, we obtain\begin{widetext}
\begin{eqnarray}
 &  & \left[\frac{1}{a_{{\rm bf}}}-\frac{\sqrt{2M(M+1)}}{M+1}\sqrt{\frac{M+2}{2(M+1)}K^{2}+\frac{1}{a_{{\rm f}}^{2}}-i\varepsilon}\right]\zeta({\bf K},\varepsilon)+\frac{(M+1)}{(2\pi)^{2}M}\int d{\bf K}'\frac{\zeta({\bf K}',\varepsilon)}{\frac{M+1}{2M}(K'^{2}+K^{2})+\frac{1}{M}{\bf K}\cdot{\bf K}'+\frac{1}{a_{{\rm f}}^{2}}-i\varepsilon}\nonumber \\
 &  & -\frac{2\pi(M+1)}{M}\int\frac{d{\bf K}'}{(2\pi)^{3}}\frac{4\pi a({\bf K}',\varepsilon)}{(K'^{2}-\frac{4M}{M+2}i\varepsilon)\left(\frac{M+1}{2M}K'^{2}+K^{2}+{\bf K}\cdot{\bf K}'+\frac{1}{a_{{\rm f}}^{2}}-i\varepsilon\right)}+\frac{2\pi(M+1)}{M(K^{2}+\frac{1}{a_{{\rm f}}^{2}}-i\varepsilon)}\nonumber \\
 &  & -\frac{2\pi(M+1)i\varepsilon}{M\left(K^{2}+\frac{1}{a_{{\rm f}}^{2}}-i\varepsilon\right)(K^{2}+\frac{1}{a_{{\rm f}}^{2}})}=0.\label{eq:-15}
\end{eqnarray}
\end{widetext}

Equations (\ref{eq:-14}) and (\ref{eq:-15}) are the integral equations
satisfied by $a({\bf K},\varepsilon)$ and $\zeta({\bf K},\varepsilon)$.
We can further express $a({\bf K},\varepsilon)$ and $\zeta({\bf K},\varepsilon)$
as $a({\bf K},\varepsilon)=\sum_{l,m}a_{l,m}(K,\varepsilon)Y_{l}^{m}(\hat{{\bf K}})$
and $\zeta({\bf K},\varepsilon)=\sum_{l,m}\zeta_{l,m}(K,\varepsilon)Y_{l}^{m}(\hat{{\bf K}})$,
where $\hat{{\bf K}}$ is the unit vector along the direction of ${\bf K}$,
and $Y_{l}^{m}$ is the spherical hormonic with degree $l$ and order
$m$. Then we can obtain equations of $a_{l,m}(K,\varepsilon)$ and
$\zeta_{l,m}(K,\varepsilon)$. It is easy to prove that the equations
for different values of $(l,m)$ are decoupled with each other. Furthermore,
the inhomogeneous term appears only in equations for $a_{0,0}(K,\varepsilon)$
and $\zeta_{0,0}(K,\varepsilon)$. As a result, we have $a_{l,m}(K,\varepsilon)=\zeta{}_{l,m}(K,\varepsilon)=0$
for $l>0$. Therefore, the solution of Eqs. (\ref{eq:-14}) and (\ref{eq:-15})
is independent of the direction of ${\bf K}$, and thus can be expressed
as 
\begin{equation}
a({\bf K},\varepsilon)=a(K,\varepsilon),\ \zeta({\bf K},\varepsilon)=\zeta(K,\varepsilon).\label{eq:-16}
\end{equation}
Using this result, Eqs. (\ref{eq:-14}) and (\ref{eq:-15}) can be
simplified as Eqs. (\ref{stm1-1}) and (\ref{stm2-1}) in our main
text. These two equations are the STM equations in our system. As
shown in the main text, in our problem the three-body boundary condition
in the region where all the three atoms are close with each other
is necessary. Such a condition is provided by the momentum cutoff $\Lambda e^{i\eta}$
in the integrations in Eqs. (\ref{stm1-1}) and (\ref{stm2-1}) \cite{pascal fran}.

\section{Single-Pole Approximation}

In this Appendix we show our calculation of atom-dimer scattering length
with single-pole approximation, and derive Eq. (\ref{spa}) in
our main text. As we have discussed before, the atom-dimer
scattering length is given by Eq. (\ref{aad-1}): 
\begin{equation}
a_{{\rm ad}}=\lim_{\varepsilon\rightarrow0^{+}}\left(\frac{8M\pi^{2}}{M+2}\right){}_{12}\langle\phi_{b}|{}_{3-12}\langle0|(V_{23}+V_{31})|\Psi_{+}(\varepsilon)\rangle,\label{eq:-17}
\end{equation}
with $|\Psi_{+}(\varepsilon)\rangle$ given by Eq. (\ref{eq:-2}):
\begin{equation}
|\Psi_{+}(\varepsilon)\rangle=|\Psi_{{\rm in}}\rangle+G_{3}(\varepsilon)(V_{23}+V_{31})|\Psi_{+}(\varepsilon)\rangle.\label{eq:-2-1}
\end{equation}
Here the Green's function $G_{3}(\varepsilon)$ is defined in Eq.
(\ref{g3}), and the states and operators are defined as before.

As shown in main text, in the single-pole approximation, we have 
\begin{equation}
G_{3}(\varepsilon)\approx\int d{\bf K}\frac{|{\bf K}\rangle_{3-12}\langle{\bf K}|\otimes|\phi_{b}\rangle_{12}\langle\phi_{b}|}{i\varepsilon-\left(\frac{M+2}{4M}\right){\bf K}^{2}}.\label{g3-1}
\end{equation}
Substituting Eq. (\ref{g3-1}) into Eq. (\ref{eq:-2-1}), we find
that under the single-pole approximation there is 
\[
|\Psi_{+}(\varepsilon)\rangle=|\phi_{b}\rangle_{12}|\psi\rangle_{3-12},
\]
where $|\psi\rangle_{3-12}$ satisfies the relation 
\begin{equation}
_{3-12}\langle{\bf K}|\psi\rangle_{3-12}=\delta({\bf K})+\frac{T({\bf K},\varepsilon)}{i\varepsilon-\left(\frac{M+2}{4M}\right){\bf K}^{2}}\label{eq:-18}
\end{equation}
with 
\begin{eqnarray}
T({\bf K},\varepsilon) & = & _{3-12}\langle{\bf K}|{}_{12}\langle\phi_{b}|(V_{23}+V_{31})|\Psi_{+}(\varepsilon)\rangle\label{eq:-19}\\
 & = & \int d{\bf K}'\left[_{3-12}\langle{\bf K}|{}_{12}\langle\phi_{b}|(V_{23}+V_{31})|\phi_{b}\rangle_{12}|{\bf K}'\rangle_{3-12}\right]\nonumber \\
 &  & \times\left(_{3-12}\langle{\bf K}'|\psi\rangle_{3-12}\right).\label{eq:-20}
\end{eqnarray}
According to Eqs. (\ref{eq:-17}) and (\ref{eq:-19}), we have 
\begin{equation}
a_{{\rm ad}}=\lim_{\varepsilon\rightarrow0^{+}}\left(\frac{8M\pi^{2}}{M+2}\right)T(0,\varepsilon).\label{eq:-21}
\end{equation}

On the other hand, by substituting Eq. (\ref{eq:-20}) into Eq. (\ref{eq:-18}),
we can obtain the integral equation for $_{3-12}\langle{\bf K}|\psi\rangle_{3-12}$,
from which an integral equation for $T({\bf K},\varepsilon)$ can
be derived. Using a similar analysis as in the paragraph before
Eq. (\ref{eq:-16}), we can find that the solution of such an equation
is independent of the direction of ${\bf K}$, i.e., we have $T({\bf K},\varepsilon)=T(K,\varepsilon)$,
and the integral equation can be simplified as in Eq. (\ref{spa}) of
our main text.



\begin{thebibliography}{10}
\bibitem{mean-field1} R. Haussmann, Z. Phys. B: Condens. Matter \textbf{91},
291 (1993).

\bibitem{mean-field2} C. A. R. Sá de Melo, M. Randeria, and J. R.
Engelbrecht, Phys. Rev. Lett. \textbf{71}, 3202 (1993).

\bibitem{few-body} D. S. Petrov, C. Salomon, and G. V. Shlyapnikov,
Phys. Rev. Lett. \textbf{93}, 090404 (2004).

\bibitem{pair_flucuation} P. Pieri and G. C. Strinati, Phys. Rev.
B \textbf{61}, 15370 (2000).

\bibitem{exp_dimer} Y. I. Shin, A. Schirotzek, C. H. Schunck, and
W. Ketterle, Phys. Rev. Lett. \textbf{101}, 070404 (2008).

\bibitem{exp} I. Ferrier-Barbut, M. Delehaye, S. Laurent, A. T. Grier,
M. Pierce, B. S. Rem, F. Chevy and C. Salomon, Science, \textbf{345},
1035 (2014).

\bibitem{Stringari} T. Ozawa, A. Recati, M. Delehaye, F. Chevy, and S. Stringari, Phys.
Rev. A \textbf{90}, 043608 (2014).

\bibitem{pu} B. Ramachandhran, S. G. Bhongale, and H. Pu, Phys. Rev.
A \textbf{83}, 033607 (2011).

\bibitem{jetp} A. F. Andreev and E. P. Bashkin, Zh. Eksp. Teor. Fiz.
\textbf{69}, 319 (1975); G. E. Volovik, V. P. Mineev, and I. M. Khalatnikov,
Zh. Eksp. Teor. Fiz. \textbf{69}, 675 (1975).

\bibitem{Cui} A recent work calculated the regime with $a_{{\rm bf}}<0$
or $a_{{\rm bf}}>a_{{\rm s}}$: X. Cui, Phys. Rev. A {\bf 90}, 041603(R) (2014).

\bibitem{petrov03} D. S. Petrov, Phys. Rev. A \textbf{67}, 010703(R)
(2003).

\bibitem{petrov05} D. S. Petrov, C. Salomon, and G. V. Shlyapnikov,
Phys. Rev. A \textbf{71}, 012708 (2005).

\bibitem{huangyang} K. Huang, C.N. Yang, Phys. Rev. \textbf{105},
767 (1957).

\bibitem{sc} J. R. Taylor, \textit{Scattering Theory} (Wiley, New
York, 1972); W. Glöckle, \emph{The Quantum Mechanical Few-Body Problem}
(Springer, New York 1983).

\bibitem{stm} G. V. Skorniakov, K. A. Ter-Martirosian, Sov. Phys.
JETP \textbf{4}, 648 (1957) .

\bibitem{supp} See our Supplementary Materials for the brief derivation
of the STM equation and the single-pole approximation.

\bibitem{ad-petrov} D. S. Petrov, Phys. Rev. Lett. \textbf{93}, 143201
(2004).

\bibitem{ad-mora} C. Mora, R. Egger, A. O. Gogolin, and A. Komnik,
Phys. Rev. Lett. \textbf{93}, 170403 (2004).

\bibitem{ad-levinsen09} J. Levinsen, T. G. Tiecke, J. T. M. Walraven,
and D. S. Petrov, Phys. Rev. Lett. \textbf{103}, 153202 (2009).

\bibitem{ad-Nakajima} S. Nakajima, M. Horikoshi, T. Mukaiyama, P.
Naidon, and M. Ueda, Phys. Rev. Lett. \textbf{105}, 023201 (2010).

\bibitem{ad-Lompe} T. Lompe, T. B. Ottenstein, F. Serwane, K. Viering,
A. N. Wenz, G. Zürn, and S. Jochim, Phys. Rev. Lett. \textbf{105},
103201 (2010).

\bibitem{ad-Hammer} H. -W. Hammer, D. Kang, and L. Platter, Phys.
Rev. A \textbf{82}, 022715 (2010).

\bibitem{ad-Alzetto10} F. Alzetto, R. Combescot, and X. Leyronas,
Phys. Rev. A \textbf{82}, 062706 (2010).

\bibitem{ad-Braaten} E. Braaten, H. -W. Hammer, D. Kang, and L. Platter,
Phys. Rev. A \textbf{81}, 013605 (2010).

\bibitem{ad-Levinsen} J. Levinsen, D. S. Petrov, Eur. Phys. J. D
\textbf{65}, 67 (2011).

\bibitem{ad-Alzetto12} F. Alzetto, R. Combescot, and X. Leyronas,
Phys. Rev. A \textbf{86}, 062708 (2012).

\bibitem{ad-Ngampruetikorn} V. Ngampruetikorn, M. M. Parish, J. Levinsen,
Eur. Phys. Lett. \textbf{102}, 13001 (2013).

\bibitem{newaad} M. Jag, M. Zaccanti, M. Cetina, R. S. Lous, F. Schreck,
R. Grimm, D. S. Petrov, and J. Levinsen, Phys. Rev. Lett. \textbf{112},
075302 (2014).







\bibitem{three-body-hutson} M. Berninger, A. Zenesini, B. Huang,
W. Harm, H. -C. Nägerl, F. Ferlaino, R. Grimm, P. S. Julienne, and
J. M. Hutson, Phys. Rev. Lett. \textbf{107}, 120401 (2011).







\bibitem{three-body-cheng} C. Chin, arXiv:1111.1484.

\bibitem{three-body-jia} J. Wang, J. P. D'Incao, B. D. Esry, and
C. H. Greene, Phys. Rev. Lett. \textbf{108}, 263001 (2012).

\bibitem{three-body-zwerger} R. Schmidt, S. P. Rath, and W. Zwerger,
Eur. Phys. J. B \textbf{85}, 386 (2012).

\bibitem{three-body-yujun} Y. Wang and P. S. Julienne, Nat. Phys. {\bf 10}, 768 (2014).




\bibitem{pascal sepe} P. Naidon, S. Endo, and M. Ueda, Phys. Rev.
Lett. \textbf{112}, 105301 (2014).

\bibitem{pascal early} P. Naidon, S. Endo, and M. Ueda, Phys. Rev.
A. \textbf{90}, 022106 (2014).

\bibitem{Rvdw-yan} Z. -C. Yan, J. F. Babb, A. Dalgarno, and G. W.
F. Drake, Phys. Rev. A \textbf{54}, 2824 (1996).

\bibitem{Rvdw-chin} C. Chin, R. Grimm, P. S. Julienne, and E. Tiesinga,
Rev. Mod. Phys. \textbf{82}, 1225 (2010).

\bibitem{Rvdw-paul} P. S. Julienne and J. M. Hutson, Phys. Rev. A
\textbf{89}, 052715 (2014).

\bibitem{eos} N. Navon, S. Nascimbene, F. Chevy, and C. Salomon,
Science \textbf{328}, 729 (2010).

\bibitem{pascal fran} P. Naidon, and M. Ueda, C. R. Phys
\textbf{12}, 13 (2011).\end{thebibliography}
\end{document}